\documentstyle[aps,twocolumn]{revtex}
\input epsf

\def\oz{|0\rangle_z}
\def\lz{|1\rangle_z}
\def\ox{|0\rangle_x}
\def\lx{|1\rangle_x}

\begin{document}

\draft

\title{Coherent eavesdropping strategies for the 4 state quantum
cryptography protocol}

\author{J.I. Cirac$^{(1,2)}$ and N. Gisin$^{(3)}$}
\address{(1) Departamento de Fisica Aplicada, Universidad de Castilla--La
Mancha,
13071 Ciudad Real, Spain}

\address{(2) Institut f{\"u}r Theoretische Physik,
Universit{\"a}t Innsbruck, A--6020
Innsbruck, Austria}

\address{(3) Group of Applied Physics, University of Geneva, 1211 Geneva,
Switzerland}

\date{\today}

\maketitle

\begin{abstract}
An elementary derivation of best eavesdropping strategies for the 4
state BB84 quantum cryptography protocol is presented, for both
incoherent and two--qubit coherent attacks. While coherent attacks do
not help Eve to obtain more information, they are more powerful to reveal 
the whole message sent by Alice. Our results are based on symmetric
eavesdropping strategies, which we show to be sufficient to analyze these 
kind of problems.
\end{abstract}


\narrowtext


In the 4-state BB84 quantum cryptography protocol \cite{BB84}, Alice,
the {\it transmitter}, sends qubits in states chosen at random among a
set of 4 possible states ${\cal S}=\{\oz,\lz,\ox,\lx\}$, where the
subscripts $z$ and $x$ denote the polarization axis. At the other end of
the quantum channel Bob, the {\it receiver}, analyzes the qubits in a
basis chosen at random between $\{\oz,\lz\}$ ($z$ basis) and
$\{\ox,\lx\}$ ($x$ basis). When Bob's basis happens to be compatible with
the state sent by Alice, the only cases that the two partners will
consider after publicly revealing the basis used for each qubit, Bob's
results are perfectly correlated to the states sent by Alice. The safety
of this protocol relies on the control of this perfect correlation and
on the quantum principle that any eavesdropping necessarily perturbs the
state of some of the qubits, hence reducing this correlation. Alice and
Bob measure this correlation by publicly comparing a sample of their
data. However, there is a practical question: given a measured
correlation, how much information could have been collected by a
malevolent third party, called traditionally Eve?

The general eavesdropping attack consists of first letting an auxiliary
quantum system, called the probe, interact with the message (in form of
qubits) sent by Alice \cite{FP96}. Then, Eve waits until she knows the
basis used in Bob's measurements, and extracts as much information as
possible out of her probe. Thus, a given eavesdropping strategy can be
characterized by a pair $({\cal U},{\cal M})$, where ${\cal U}$ is the
unitary transformation used in the interaction between Eve's probe and
the message, and ${\cal M}$ represents Eve's measurement. Most of the
analysis of the best eavesdropping strategy so far have been
concentrated in the so--called ``incoherent attacks'', in which each of
Eve's probes interacts independently with each qubit sent by Alice.
However, there might exist a better strategy for Eve in which her probes
interact with more than one of Alice's qubits at the same time, what is
known as ``coherent attacks''. This letter provides an elementary
derivation of the optimal eavesdropping strategy for a given disturbance
of the perfect correlation between Alice and Bob's data for both
incoherent attacks and coherent attacks on two qubits. Our results show
that coherent attacks do not help Eve to obtain more information about
Alice message. However, they allow Eve to obtain the whole message sent
by Alice with a higher probability. Thus, this is a situation where
coherent attacks prove to be more powerful than incoherent (single
qubit) ones \cite{BFS} even if no error correction or privacy amplification
\cite{Privamp} is performed.

Our formulation of the problem is based on what we will call ``symmetric
eavesdropping strategies''. Those are attacks in which Eve treats all
the possible messages sent by Alice on an equal footing. As we show
below, without loss of generality, one can restrict oneself to analyze
symmetric eavesdropping strategies only, which reduces enormously the
problem. The reason is that for any given non--symmetric attack,
coherent or incoherent, one can always find a symmetric one that
reproduces the results of the non--symmetric strategy. 

This paper is organized as follows: first, we consider general symmetric
incoherent attacks; next, we consider general symmetric 2-qubit coherent
attacks; and finally, we show that the symmetric strategies are
sufficient to study eavesdropping attacks. 

Let us analyze first the incoherent attacks. Denoting by $|E\rangle$ the
initial state of Eve's probe, the unitary operation ${\cal U}$ can be
characterized by its action on a basis set of Alice states
\begin{mathletters}
\label{Ez}
\begin{eqnarray}
|E\rangle &\otimes& \oz \stackrel{\cal U}{\longrightarrow} 
  |E_{0,0}^z\rangle \oz + |E_{0,1}^z\rangle \lz,\\
|E\rangle &\otimes& \lz \stackrel{\cal U}{\longrightarrow} 
  |E_{1,0}^z\rangle \oz + |E_{1,1}^z\rangle \lz,
\end{eqnarray}
\end{mathletters}
where $|E_{i,j}^z\rangle$ are unnormalized states of Eve's probe that
span a 4--dimensional Hilbert space. Equation (\ref{Ez}) can be written
in a more compact form as 
\begin{equation}
\label{Ez1}
|E\rangle \otimes
\left(\begin{array}{c} \oz \\ \lz \end{array} \right) 
\stackrel{\cal U}{\longrightarrow} 
{\cal E}^z \otimes
\left(\begin{array}{c} \oz \\ \lz \end{array} \right),
\end{equation}
{w}here ${\cal E}^z$ is a $2\times 2$ matrix whose elements are Eve's
states $|E_{i,j}^z\rangle$. The action of ${\cal U}$ on the basis 
$\{\ox,\lx\}$ can be readily derived using linearity
\begin{equation}
\label{Ex2}
|E\rangle \otimes
\left(\begin{array}{c} \ox \\ \lx \end{array}\right)
\stackrel{\cal U}{\longrightarrow}
{\cal E}^x \otimes
\left(\begin{array}{c} \ox \\ \lx \end{array} \right).
\end{equation}
Here ${\cal E}^x = {\cal V} {\cal E}^z {\cal V}^\dagger$, where
\begin{equation}
\label{V}
{\cal V} = {\cal V}^\dagger= \frac{1}{\sqrt{2}}
\left(\begin{array}{rr} 1 & 1 \\ 1 & -1 \end{array}\right)
\end{equation}
implements the transformation from the $z$ to the $x$ basis.

Let us now particularize the above equations for a symmetric attack. To
this aim, we note that for a given measurement strategy ${\cal M}$, the
scalar products $\langle E_{i,j}^z| E_{k,l}^z\rangle$ characterize the
unitary operation ${\cal U}$ of Eve's attack. In the present context, a
symmetric attack is defined by imposing that all scalar products
$\langle E_{i,j}^\alpha| E_{k,l}^\alpha\rangle$ ($\alpha=x,z$,
$i,j,k,l=0,1$)) have to be invariant with respect to two operations: (i)
exchange of the states $0 \leftrightarrow 1$; (ii) exchange of basis
$z\leftrightarrow x$. This amounts to requiring that all possible states
sent by Alice ($\in {\cal S}$) are treated on an equal footing. For
example, the first condition (i) means that the probability amplitudes
of any outcome should not depend on whether Alice prepared a $|0\rangle$
or a $|1\rangle$. It imposes that the scalar products should be
invariant under the exchange of the subindices $0 \leftrightarrow 1$, i.e.,
\begin{mathletters}
\label{cond1}
\begin{eqnarray}
F &=& \langle E_{0,0}| E_{0,0} \rangle = \langle E_{1,1}| E_{1,1} \rangle,\\
D &=& \langle E_{0,1}| E_{0,1} \rangle = \langle E_{1,0}| E_{1,0} \rangle,\\
F_1 &=& \langle E_{0,0}| E_{1,1} \rangle = \langle E_{1,1}| E_{0,0} \rangle,\\
D_1 &=& \langle E_{0,1}| E_{1,0} \rangle = \langle E_{1,0}| E_{0,1} \rangle,
\end{eqnarray}
\end{mathletters}
etc, where traditionally $F$ is called fidelity and $D$ disturbance. 
The second condition (ii) means that the outcomes should not depend on
whether Alice used the $z$ or the $x$ basis. Together with the normalization
condition, it imposes that
\begin{equation}
\label{relations}
F + D = 1, \quad\quad F - D = F_1 + D_1,
\end{equation}
and that all the other scalar products that do not appear in
(\ref{cond1}) have to be zero. Thus, the number of independent
parameters in the problem is reduced to only 2 real numbers ($D$ and
$D_1$, for example). Fixing the disturbance $D$, one can then determine
any property of the attack (information gain, probability of success in
guessing the qubit sent by Alice, etc) as a function of 1 parameter
only. Therefore, best strategies can be very easily (even analytically) found
by maximizing functions with respect to one parameter. 

With these notations one can rewrite Eqs.~(\ref{Ez}) in the following
appealing form:
\begin{equation}
|E\rangle \otimes \oz \stackrel{\cal U}{\longrightarrow} 
  \sqrt{F}|\hat E_{0,0}^z\rangle \oz + \sqrt{D}|\hat E_{0,1}^z\rangle \lz
\end{equation}
where $|\hat E_{j,k}^z\rangle$ represent Eve's (normalized) state in
case Alice send a $j$ bit and Bob detected a $k$ bit. The fidelity $F$
is the probability that Bob detected Alice bit correctly, while the
disturbance $D$ is the complementary probability of wrong detection.

Let us analyze now the measurement that allows Eve to extract the
maximal information out of her probe. In principle, one should consider
separately the cases in which Alice prepares each of the states ${\cal
S}$, and then average over all these cases. However, thanks to the
symmetry, it is sufficient to consider only the situation in which Alice
chooses the $z$ basis. In that case, after tracing over Alice's
particle, Eve's two possible states are:
\begin{mathletters}
\begin{eqnarray}
\rho_E(0) &=& |E_{0,0}^z \rangle \langle E_{0,0}^z| + 
  |E_{0,1}^z \rangle \langle E_{0,1}^z|,\\
\rho_E(1) &=& |E_{1,1}^z \rangle \langle E_{1,1}^z| + 
  |E_{1,0}^z \rangle \langle E_{1,0}^z|,
\end{eqnarray}
\end{mathletters}
The optimal measurement strategy ${\cal M}$ for Eve is based on the
fact that the four states that appear in the decomposition of these two
density matrices, $|E_{i,j}^z\rangle$, fall into two mutually orthogonal sets,
$S_0=\{|E_{0,0}^z\rangle,|E_{1,1}^z\rangle\}$ and
$S_1=\{|E_{0,1}^z\rangle,|E_{1,0}^z\rangle\}$, that can be therefore
distinguished deterministically. The first set $S_0$ occurs with probability
$F$, in which case Eve has to extract information about two states with
same a priori probability and overlap $\cos(\alpha)\equiv F_1/F$. The
second set $S_1$ occurs with probability $D$ and state overlap
$\cos(\beta)\equiv D_1/D$ \cite{footnote}. Consequently, Eve's information
equals:
\begin{equation}
\label{IE}
I_e = 1 + F \; H(P_\alpha) + D \; H(P_\beta),
\end{equation}
where $H(P)=P\ln(P)+(1-P)\ln(1-P)$ is the Shanon entropy, and $P_\alpha=
[1+\sin(\alpha)]/2$. The probability that Eve finds out the state sent
by Alice is 
\begin{equation}
P_e = F \;P_\alpha + D \;P_\beta.
\end{equation}
These two quantities $I_e$ and $P_e$ are maximum for $\alpha=\beta$ (or,
equivalently, $F_1/F=D_1/D$). In this case one has
$D=[1-\cos(\alpha)]/2$, $I_e^{\rm opt}=1+H(P_\alpha)$, and $P_e^{\rm
opt}=P_\alpha$. These relations provide the explicit and analytic
optimum eavesdropping strategy with a single parameter $\alpha$. It
agrees with the numerical result and the bound devised in
\cite{privatecom}. A complete presentation can be found in \cite{FGGNP}.
In Fig.~1 (upper plot) we have plotted $I_e^{\rm opt}$, Bob's
information $I_b$, and $I_e^{\rm opt}+I_b$ as a function of $D$. In the
lower plot we represent $P_e^{\rm opt}$ and Bob's probability $P_b=F$.
For disturbances below 5\%, as happens in experiments
\cite{Townsend94,MullerNature95,Franson95}, the information gain is
proportional to the disturbance: $I_{e}\approx2\ln(2)D$. In this case a
simpler 2--dimensional probe suffices \cite{GH}. On the other extreme,
for $\alpha=\pi/3$ the disturbance is ${1\over 4}$ and the optimal
information gain is $0.6454$ ($0.579$ for 2-dimensional probe).

Let us conclude the analysis of incoherent attacks with the intriguing
case when Bob's information $I_B=1+H(D)$ (probability of success
$P_B=1-D$) equals Eve's optimum information $I_e^{\rm opt}$ ($P_e^{\rm
opt}$). This occurs when $\cos(\alpha)=\sin(\alpha)$, i.e. when
$\alpha=\pi/4$. Remarkably this coincides precisely with the threshold
of violation of the Bell-CHSH inequality \cite{CHSH} by Alice and Bob.
Indeed, the S parameter whose value is restricted below 2 by local
realism, $S=2\sqrt{2}\cos(\alpha)=2$ for $\alpha=\pi/4$. This proves a
conjecture presented in \cite{GH}. Another intriguing observation is
that $I_e^{\rm opt} + I_b \le 1$ (see Fig.~1); that is, the sum of the
informations gained by Eve and Bob does not add to 1. This is due to the
fact that Eve and Bob share an entangled state, and perform only local
measurements. Note however, that if Bob tells Eve the outcome of his
measurement, then Eve can guess with certainty what was the qubit sent
by Alice. In other words, if Bob and Eve are allowed to communicate one
with each other, then, despite they do not perform joint measurements of
the entangled state, they can determine the state sent by Alice.

\begin{figure}[ht]
\begin{center}
   \hspace{0mm}
    \epsfxsize=6.5cm
    \epsfbox{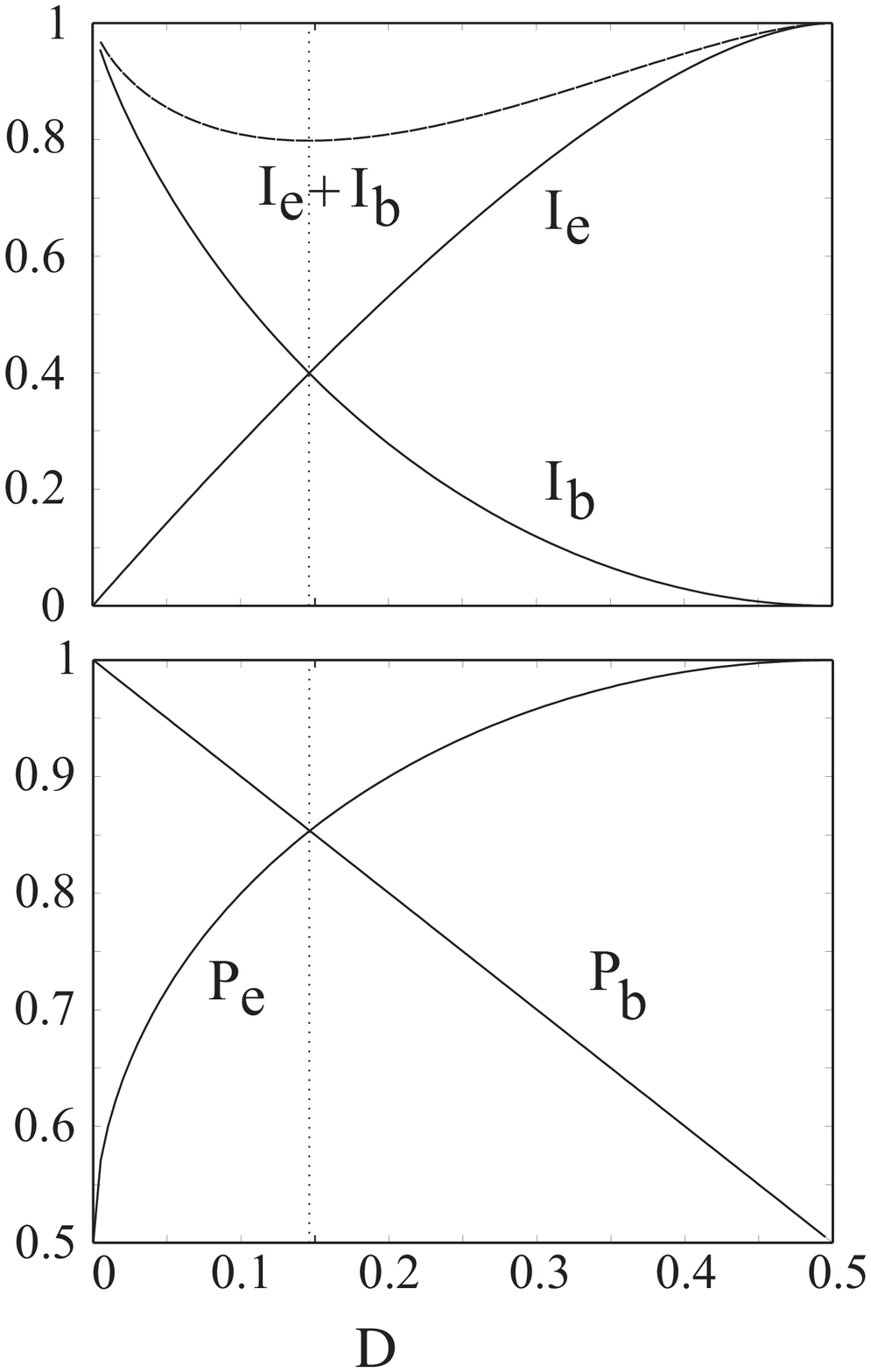}\\[0.2cm]
\begin{caption}
{\sf
Upper plot: Eve's optimum information $I_e^{\rm opt}$, Bob's information 
$I_b$, and $I_e^{\rm opt}+I_b$ (dashed line) as a function of
the disturbance $D$. Lower plot: Eve's ($P_e$) and Bob's ($P_b$) 
probability of obtaining the message sent by Alice as a function
of $D$.
}
\end{caption}
\end{center}
\end{figure}

We~ analyze now the possibility for Eve to interact coherently with two
qubits sent by Alice. We consider that Alice selects one of the 16
possible states among the set ${\cal
S}=\{\oz\oz,\oz\lz,\lz\oz,\lz\lz,\ox\oz,\ox\lz,\ldots\}$. For
convenience, we use binary notation: $|0\rangle_{zx} \equiv \oz\ox$,
$|1\rangle_{xz}\equiv \lx\oz$, $|2\rangle_{zz}\equiv \oz\lz$,
$|3\rangle_{xx}\equiv \lx\lx$, etc. Eve's unitary operation is
characterized by the action of ${\cal U}$ on the basis $zz$
\begin{equation}
\label{Ez2}
|E\rangle \otimes
\left(\begin{array}{c} 
|0\rangle_{zz} \\ |1\rangle_{zz} \\ |2\rangle_{zz} \\ |3\rangle_{zz} \\
\end{array}\right)
\stackrel{\cal U}{\longrightarrow}
{\cal E}^{zz} \otimes
\left(\begin{array}{c}
|0\rangle_{zz} \\ |1\rangle_{zz} \\ |2\rangle_{zz} \\ |3\rangle_{zz} \\
\end{array} \right),
\end{equation}
where ${\cal E}$ is a $4\times 4$ matrix containing Eve's probe states,
i.e., 
\begin{equation}
\label{Ez3}
{\cal U}|E\rangle\otimes |i\rangle_{zz} = \sum_{j=0}^3 |E_{i,j}^{zz}\rangle
\otimes |j\rangle_{zz},\quad\quad (i=0,\ldots,3).
\end{equation} 
Taking into account the unitarity of ${\cal U}$, this operation is
characterized by 240 real numbers (which are directly related to the
scalar products of the elements of ${\cal E}^{zz}$). By imposing 
the conditions of a symmetric attack, and without loss of generality,
the number of independent parameters is reduced to only 5.

The symmetric attack imposes now three conditions for the scalar
products $\langle E_{i,j}^{zz}| E_{k,l}^{zz} \rangle$, which are
analogous to the ones corresponding to the incoherent attack. They have to
remain unchanged under: (i) the exchange $|0\rangle \leftrightarrow
|1\rangle$ in the first qubit {\it or} the second qubit (independently);
(ii) the exchange of the state of the first qubit by the state of the
second qubit; (iii) the change of the basis (i.e., $zz$ to $xz$ or $zx$
or $xx$). The first condition is equivalent to demanding that the
scalar products have to remain unchanged if one exchanges all the
subindices 
\begin{mathletters}
\label{subst}
\begin{equation}
\label{subst1}
1\leftrightarrow 2
\end{equation}
The second condition implies invariance if one exchanges simultaneously the
subindices 
\begin{equation}
\label{subst2}
0\leftrightarrow 1 \quad {\rm and} \quad 2\leftrightarrow 3. 
\end{equation}
\end{mathletters}
Once this is satisfied, the third condition amounts to simply impose that
$\langle E_{i,j}^{zz}| E_{k,l}^{zz}\rangle = \langle E_{i,j}^{xz}|
E_{k,l}^{xz}\rangle, \forall i,j,k,l=0,\ldots,3$, where the matrix
${\cal E}^{xz} = ({\cal V}\otimes 1) {\cal E}^{zz} 
({\cal V}^\dagger\otimes 1)$, 
and ${\cal V}$ is defined in (\ref{V}) \cite{footnote2}. In more detail,
let us define the following parameters:
\begin{eqnarray*}
&A=\langle E_{0,0}^{zz}| E_{0,0}^{zz}\rangle, \quad
 A_1=\langle E_{0,0}^{zz}| E_{1,1}^{zz}\rangle, \quad
 A_2=\langle E_{0,0}^{zz}| E_{3,3}^{zz}\rangle,& \\
&B=\langle E_{0,1}^{zz}| E_{0,1}^{zz}\rangle, \quad\quad\quad
  B_1=\langle E_{0,1}^{zz}| E_{1,0}^{zz}\rangle,& \\  
&B_2=\langle E_{0,1}^{zz}| E_{3,2}^{zz}\rangle, \quad\quad\quad  
 B_3=\langle E_{0,1}^{zz}| E_{2,3}^{zz}\rangle,& \\
&C=\langle E_{0,3}^{zz}| E_{0,3}^{zz}\rangle, \quad
 C_1=\langle E_{0,3}^{zz}| E_{1,2}^{zz}\rangle, \quad
 C_2=\langle E_{0,3}^{zz}| E_{3,0}^{zz}\rangle.&
\end{eqnarray*}
Imposing the symmetry and unitary conditions one finds that: 
{\it first}, the scalar product
$\langle E_{i,j}^{zz}| E_{k,l}^{zz}\rangle$ is zero if
among the indices $(i,j,k,l)$ either three are identical
and one is different or two are idential and two different \cite{footnote3};
{\it second}, each of the other scalar products are identical to one
of these parameters \cite{footnote4}; {\it third}, all these parameters are real
and fulfill the following relations [analogous to (\ref{relations})]
\begin{eqnarray}
\label{relations2}
A+2B+C=1,&\quad\quad& B-C=B_3+C_1, \nonumber\\
A-B=A_1+B_1,&\quad\quad& A_1-A_2=B_2+B_3, \nonumber\\
B_1-C_2=B_2+C_1&&
\end{eqnarray}
In this way, Eve's unitary transformation is fully characterized
by only 5 real parameters.

By virtue of the symmetry, to determine Eve's disturbance we
can consider the case in which Alice prepared the state $|0\rangle_{zz}$.
One readily finds $D=1-(A+B)$. To determine
Eve's gain of information, we can restrict ourselves to the case
in which Alice used the $zz$ basis to prepare the states. The density
matrices that Eve obtains for the 4 different choices of Alice
qubits ($|0\rangle_{zz},\ldots,|3\rangle_{zz}$) can be directly
read off from the Eq.~(\ref{Ez3}). Analogous to the incoherent attack
case, Eve has first to distinguish among 4 orthogonal sets:
\begin{mathletters}
\begin{eqnarray} 
S_0&=&\{|E_{0,0}^{zz}\rangle,|E_{1,1}^{zz}\rangle,|E_{2,2}^{zz}\rangle,
|E_{3,3}^{zz}\rangle\},\\
S_1&=&\{|E_{0,1}^{zz}\rangle,|E_{1,0}^{zz}\rangle,|E_{2,3}^{zz}\rangle,
|E_{3,2}^{zz}\rangle\},\\
S_2&=&\{|E_{0,2}^{zz}\rangle,|E_{2,0}^{zz}\rangle,|E_{1,3}^{zz}\rangle,
|E_{3,1}^{zz}\rangle\},\\
S_3&=&\{|E_{0,3}^{zz}\rangle,|E_{3,0}^{zz}\rangle,|E_{1,2}^{zz}\rangle,
|E_{2,1}^{zz}\rangle\},
\end{eqnarray}
\end{mathletters}
{w}hich occur with probabilities $A$, $B$, $B$, and $C$, respectively.
This distinction can be performed with certainty, since the sets are
mutually orthogonal. Then she has to find out which element of 
the corresponding set she has. To analyze that, note that
in the case of $S_0$ and $S_3$, the four vectors form a pyramid with
square base in a 4-dimensional Hilbert space [Fig.~2(a)], whereas in the
case of $S_1$ and $S_2$, they form a pyramid with rectangular base
[Fig.~2(b)]. For $S_0$ and $S_3$ there are two
different overlaps among the four vectors, namely $\cos(\theta^0_{1,2})=
A_1/A,A_2/A$, and $\cos(\theta_{1,2}^3)=C_1/C,C_2/C$, respectively. 
For $S_1$ and $S_2$ there are three different overlaps
$\cos(\theta^1_{1,2,3})=\cos(\theta^2_{1,2,3})= B_1/B,B_2/B,B_3/B$.
Denoting by $\vec a_{i}^j$ ($i,j=0,\ldots,3$) the i--th vector of the
j--th set, we can then write the 4 vectors of any given set $j$ in terms
of an orthonormal (cartesian) basis $\vec e_0,\ldots,\vec e_3$ as
\begin{equation}
\left(\begin{array}{c}
\vec a_0^j \\ \vec a_1^j \\ \vec a_2^j \\ \vec a_3^j 
\end{array}\right) = 
\left(\begin{array}{cccc}
a & b & c & d \\
b & a & d & c \\
c & d & a & b \\
d & c & b & a 
\end{array}\right)
\left(\begin{array}{c}
\vec e_0^j \\ \vec e_1^j \\ \vec e_2^j \\ \vec e_3^j 
\end{array}\right).
\end{equation}
The coefficients satisfy the following equations:
\begin{mathletters}
\label{meas}
\begin{eqnarray}
a^2+b^2+c^2+d^2=1, &\quad\quad& 2(ab+cd)=cos(\theta^j_1),\\
2(ad+bc)=cos(\theta^j_2) &\quad\quad& 2(ac+bd)=cos(\theta^j_3),
\end{eqnarray}
\end{mathletters}
where we choose the solutions with $a > b,c,d$. Here, for the sake of
compactness we have defined $\theta_3^{0,3}=\theta_1^{0,3}$. We find
that the best measurement corresponds to use the cartesian basis.
For example, if the result is $\vec e_2$, then take as a guess the
corresponding state $\vec a_2$.

\begin{figure}[ht]
\begin{center}
\hspace*{0mm}
  \epsfxsize=6.5cm
  \epsfbox{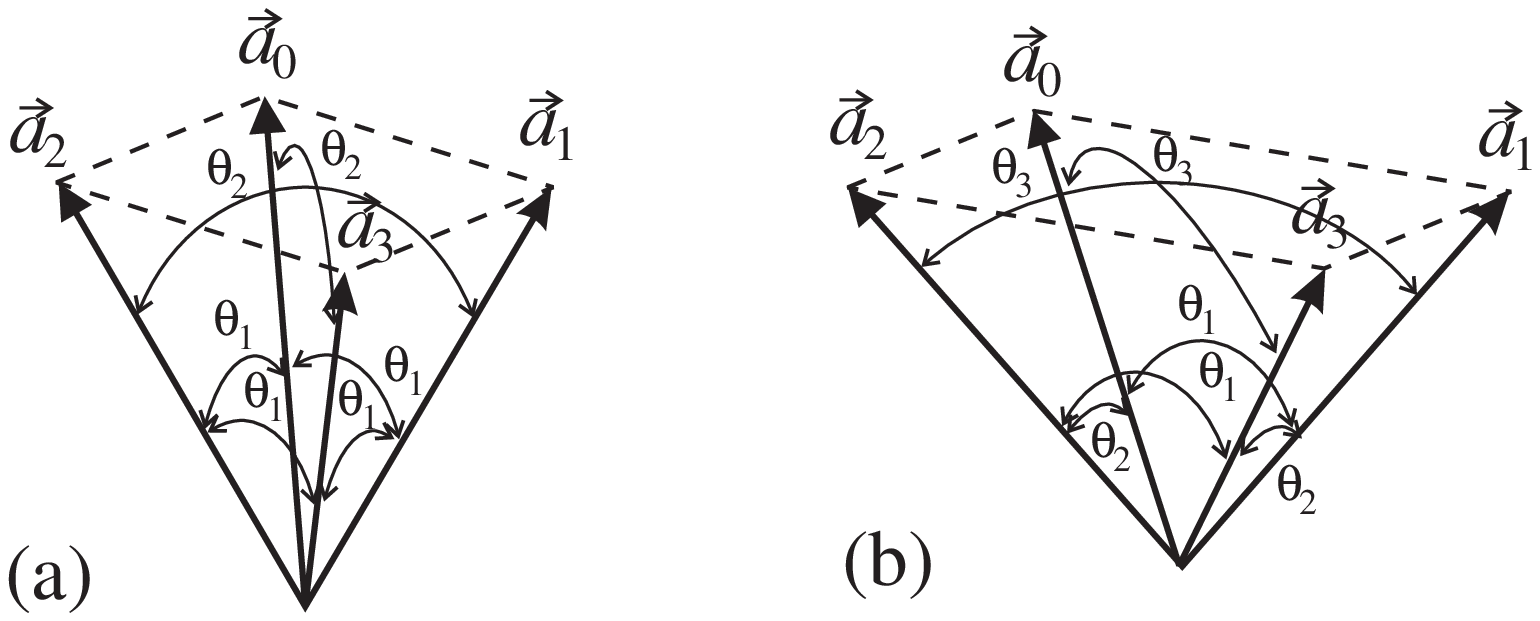}\\[0.2cm]
\begin{caption}
{\sf
Diagrammatic representation of the states of: (a) sets $S_0$ and
$S_3$; (b) sets $S_1$ and $S_2$.
}
\end{caption}
\end{center}
\end{figure}

For a given set of 5 independent parameters using the relations
(\ref{relations2})
one first calculates the
angles $\theta$, as well as the
probabilities of ending up with a given set $S$. Then one calculates the
probabilities $a^2$, $b^2$, $c^2$ and $d^2$ for each given set
by solving Eqs.~(\ref{meas}). Starting from these
probabilities, one can determine for a given disturbance $D$ the
probability for Eve to find out what was the message sent by Alice, as
well as her information gain. We have found analytically that the
strategy that gives Eve the maximum amount of information coincides with
the incoherent attack analyzed above (Fig.~1). In contrast, the strategy that
gives her maximum probability for determining the full message ($0,1,2$,
or $3$) sent by Alice is different, and indeed relies on the use of a
coherent attack.

If Fig.~3 we have plotted Eve's and Bob's probabilities $P_e$ and $P_b$
for guessing the two qubits sent by Alice as a function of the
disturbance, both for the optimum incoherent attack $P^1$ as well as for
the optimum coherent attack $P^2$ (this last, calculated numerically
with a maximization procedure). In the lower insert we have plotted a
detail of the curves around the disturbance which corresponds to
violation of Bell's inequalities in the incoherent (single qubit)
attack. In the upper insert we have plotted the relative gain of
probability that Eve and Bob obtain using the coherent attack. This
relative gain is rather small (smaller than 1.5\%). For Eve it occurs at
low disturbances, just where experiments take place. 

\begin{figure}[ht]
\begin{center}
\hspace*{0mm}
  \epsfxsize=6.5cm
  \epsfbox{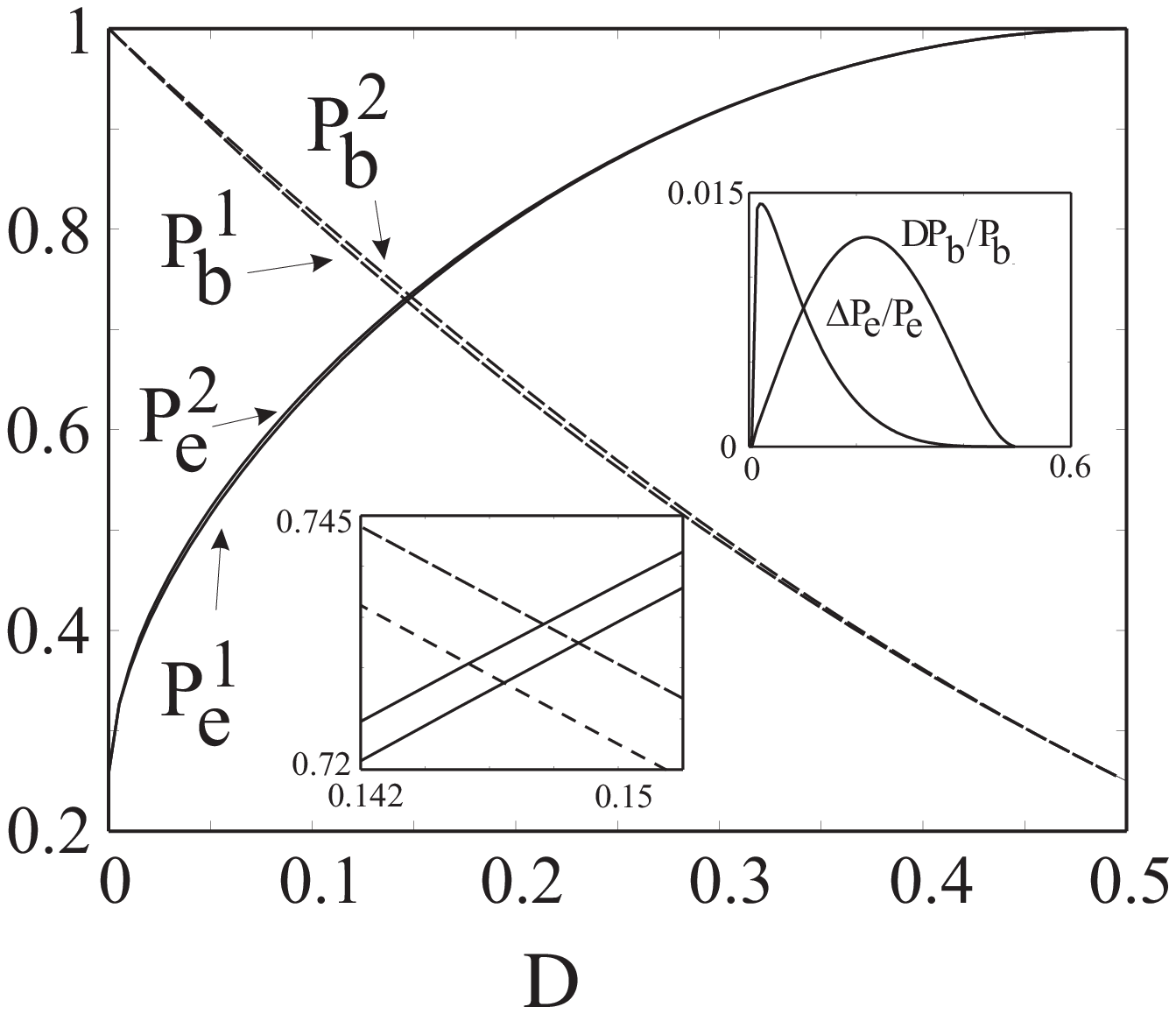}\\[0.2cm]
\begin{caption}
{\sf
Bob's ($P_b$) and Eve's ($P_e$) probabilities of guessing the
two qubits sent by Alice for incoherent ($P_1$) and coherent
($P_2$) attacks. The inserts show a detail of the curves, as
well as the relative gain of the coherent attack.
}
\end{caption}
\end{center}
\end{figure}

Let us notice that Bob could detect coherent attacks on successive pairs
of qubit by controling the correlation of failures of successive qubits.
However, Eve can prevent such a defense by attacking randomly chosen
pairs of qubits, though at the cost of extra practical complications.

Let us conclude by sketching the proof that for a general case of
coherent attack on $n$ qubits ($n\ge 1$), the analysis of what we have
defined as symmetric strategies suffices. To do that we show that for
any given strategy $\{{\cal U},{\cal M}\}$ one can always construct
a symmetric one $\{{\cal U}^s,{\cal M}^s\}$ that reproduces the results 
for all the properties which are defined as averages over the set ${\cal
S}$ of possible states sent by Alice
\begin{equation}
\label{Quant}
\overline{Q} = \frac{1}{2^{2n}} \sum_{|\alpha\rangle \in {\cal S}}
Q_\alpha.
\end{equation}
For example if $Q$ is the probability for Eve to guess the result sent
by Alice, then $Q_\alpha$ would be this probability provided Alice has
prepared the state $|\alpha\rangle$. $Q$ could also be Eve's or Bob's
information gain, the disturbance, etc. 

The action of ${\cal U}$ is given, in general, by
\begin{equation}
{\cal U}|E\rangle\otimes |i\rangle_z = \sum_{j=0}^{2^n-1} |E_{i,j}\rangle
\otimes |j\rangle_{z},\quad\quad (i=0,\ldots,2^n-1),
\end{equation} 
i.e., it is characterized by a matrix ${\cal E}$ with the states of
Eve's probe [as in (\ref{Ez2})]. Let us denote by ${\cal T}$ one of the
operations (acting on Alice's qubits) that we have used to define the
symmetric attack. To be more specific, we take ${\cal T}$ to be an
operation that: (1) Is idempotent
\begin{mathletters}
\begin{equation}
\label{c1}
{\cal T}^2=1; 
\end{equation}
(2) Leaves Alice set of states invariant 
\begin{equation}
\label{c2}
{\cal T}{\cal S}={\cal S}.
\end{equation}
\end{mathletters}
For example,the operation corresponding to exchanging $0\leftrightarrow 1$ 
in the $q$--th qubit (in the $z$ basis) would be
\begin{equation}
{\cal T}|0\rangle_q = |1\rangle_q, \quad\quad {\cal T}|1\rangle_q = |0\rangle_q,
\end{equation}
whereas the one corresponding to exchanging the basis $z\leftrightarrow x$
would be ${\cal T}\equiv{\cal V}$ defined in (\ref{V}). Note that under 
these operations, ${\cal E}$ becomes ${\cal T}{\cal E}{\cal T}$.

We now construct explicitely a symmetric attack $\{{\cal U}^{\cal T},
{\cal M}^{\cal T}\}$ with respect to ${\cal T}$ \cite{note5}: (i) The
action of ${\cal U}^{\cal T}$ is given by the matrix ${\cal E}^{\cal T}
= \frac{1}{\sqrt{2}}({\cal E}|0\rangle_e + {\cal T}{\cal E}{\cal
T}|1\rangle_e)$, where $|0\rangle_e$ and $|1\rangle_e$ are two
orthogonal states of an extra ancilla used by Eve and ${\cal
E}|0\rangle_e$ is the matrix whose elements are the vectors obtained by
applying the tensor product to the elements of ${\cal E}$ and the vector
$|0\rangle_e$; (ii) Eve's measurement ${\cal M}^{\cal T}$ is defined as
follows: first measure the state of the ancilla; if the result is
$|0\rangle_e$, then use the original measurement strategy ${\cal M}$; if
it is $|1\rangle_e$, then use the measurement strategy ${\cal M}$ that
she would have applied if Alice state was prepared in the state ${\cal
T}|\alpha\rangle$ instead of $|\alpha\rangle$. For example, for ${\cal
T}$ corresponding to the change of basis $z\leftrightarrow x$, if Bob
announces publically a measurement along the $z$ axis and Eve obtains
$|1\rangle_e$, then she should measure her probe as if Bob had announced
a measurement in the $x$ basis. Using (\ref{c1}) one can readily check
that $\{{\cal U}^{\cal T},{\cal M}^{\cal T}\}$ fulfills $\langle{\cal
E}^{\cal T}_{i,j}|{\cal E}^{\cal T}_{k,l}\rangle = \langle({\cal T}{\cal
E}^{\cal T}{\cal T})_{i,j}| ({\cal T}{\cal E}^{\cal T}{\cal
T})_{k,l}\rangle$, i.e. it is symmetric with respect to ${\cal T}$. On
the other hand, using (\ref{c2}) one obtains that regardless of the
result of the measurement on the ancilla, ${\cal M}^{\cal T}$ will give
for $Q$ the same result as the one given by the non--symmetric strategy
(\ref{Quant}). Using this procedure, one can then construct an attack
$\{{\cal U}^s,{\cal M}^s\}$ that is symmetric with respect to all such
operations ${\cal T}$, which it is what we have called a ``symmetric
eavesdropping strategy''.

In conclusion, an analytic derivation of the incoherent eavesdropping
strategy on the BB84 4 state quantum cryptography protocol that
maximizes Eve's information has been presented. Using coherent attacks,
i.e. assuming that Eve can interact coherently with more than one of the
qubits send by Alice, seems not to improve Eve's information gain. This
was proven for the 2-qubit case and we conjecture that this result hold
for arbitrary coherent attacks. However, our analysis shows that
coherent attacks can improve the probability that Eve guesses correctly
the entire key, at the cost of higher probability that Eve guess is
wrong on each qubit. This provides another example \cite{BFS,Popescu}
that coherent (sometimes called collective) measurements can provide
results unachievable with only incoherent measurements.\\

Helpful discussions with B. Huttner, N. Lutkenthaus, and P. Zoller are
acknowledged. NG acknowledges financial support by the Swiss National
Science Foundation. This work has been supported by the TMR newtork
ERB--FMRX--CT96--0087.



\begin{references}

\bibitem{BB84} C. H. Bennett and G. Brassard, in {\it Proceedings of
the IEEE International Conference on Computer, Systems, and Signal
Processing, Bangalore, India\/} (IEEE, New York, 1984), pp. 175--179.

\bibitem{FP96} C. A. Fuchs and A. Peres, Phys. Rev. A {\bf 53}, 2038, 1996.

\bibitem{BFS}
C. Bennet, C. Fuchs, and J. Smolin, preprint (quant--ph/9611006).

\bibitem{Privamp} 
C. H. Bennett, F. Bessette, G. Brassard, L. Salvail,
and J. Smolin, J. Crypto {\bf 5}, 3, 1992.

\bibitem{footnote}
Note that at no additional cost (i.e.,
no additional disturbance), Eve knows whether Bob and Alice bits are
identical (first set) or differ (second set). Therefore Eve's possesses
the same information about the state prepared by Alice, as about the state
received by Bob.

\bibitem{privatecom}
Ch. Fuchs, A. Peres and R.B. Griffiths, private communications.

\bibitem{FGGNP} C. Fuchs, N. Gisin, R.B. Griffiths, C.S. Niu and A. Peres, 
{\it Optimal eavesdropping in quantum cryptography}, submitted to
Phys. Rev. A, 1997.

\bibitem{Townsend94} P. D. Townsend, Electron. Letts. {\bf 30}, 809, 1993.

\bibitem{MullerNature95} A. Muller, H. Zbinden and N. Gisin, Nature {\bf
378}, 449, 1995.

\bibitem{Franson95} J. D.Franson and B. C. Jacobs, Electron. Letts. {\bf
31}, 232, 1995.

\bibitem{GH} N. Gisin and B. Huttner, ``Quantum cloning,
eavesdropping, and Bell's inequality'' (preprint, quant-ph/9611041), 
Phys. Lett. A, in press, 1997.

\bibitem{CHSH}
J.F.~Clauser, M.A.~Horne, A.~Shimony and R.A. Holt, Phys. Rev. Lett. {\bf
23}, 880, 1969.


\bibitem{footnote2}
Due to the conditions (i) and (ii) one has actually to impose
this only to a few elements.

\bibitem{footnote3}
For example, $\langle E_{0,2}^{zz}| E_{1,1}^{zz}\rangle=
\langle E_{1,2}^{zz}| E_{1,3}^{zz}\rangle=
\langle E_{0,2}^{zz}| E_{2,2}^{zz}\rangle=0$;

\bibitem{footnote4}
For example, using (\ref{subst1}) one finds 
$B_1=\langle E_{0,1}^{zz}| E_{1,0}^{zz}\rangle=
\langle E_{0,2}^{zz}| E_{2,0}^{zz}\rangle$. Using
(\ref{subst2}) one further finds 
$B_1=\langle E_{2,3}^{zz}| E_{3,2}^{zz}\rangle=
\langle E_{1,3}^{zz}| E_{3,1}^{zz}\rangle$.

\bibitem{note5}
By symmetric attack with respect to a given operation ${\cal T}$ we mean that
all the scalar products of the elements of ${\cal E}$ have to
be invariant if one substitutes Alice state $|\alpha\rangle$ by 
${\cal T}|\alpha\rangle$

\bibitem{popescu}
S. Massar and S. Popescu, Phys. Rev. Lett. {\bf 74}, 1259 (1995).

\end{references}
\end{document}